\documentclass{article}
\usepackage{arxiv}
\pdfoutput=1
\usepackage[utf8]{inputenc} 
\usepackage[T1]{fontenc}    
\usepackage{hyperref}       
\usepackage{url}            
\usepackage{amsfonts}       
\usepackage{nicefrac}       
\usepackage{microtype}      
\usepackage{lipsum}
\usepackage{graphicx}
\usepackage{subfigure}
\usepackage{natbib}
\usepackage{algorithm}
\usepackage{algpseudocode}
\usepackage{cleveref}
\usepackage{multirow}

\title{De-anonymization Attacks on Neuroimaging Datasets}

\author{
  Vikram Ravindra\\
  Department of Computer Science\\
  Purdue University\\
  West Lafayette, IN 47906 \\
  \texttt{ravindrv@purdue.edu} \\
   \And
  Ananth Grama\\
  Department of Computer Science\\
  Purdue University\\
  West Lafayette, IN 47906 \\
  \texttt{ayg@cs.purdue.edu} \\
}

\begin{document}
\maketitle

\begin{abstract}
Advances in imaging technologies, combined with inexpensive storage, have led to an explosion in the volume of publicly available neuroimaging datasets. Effective analyses of these images hold the potential for uncovering mechanisms that govern functioning of the human brain, and understanding various neurological diseases and disorders. The potential significance of these studies notwithstanding, a growing concern relates to the protection of privacy and confidentiality of subjects who participate in these studies. In this paper, we present a de-anonymization attack rooted in the innate uniqueness of the structure and function of the human brain. We show that the attack reveals not only the identity of an individual, but also the task they are performing, and their efficacy in performing the tasks. Our attack relies on novel matrix analyses techniques that are used to extract discriminating features in neuroimages. These features correspond to individual-specific signatures that can be matched across datasets to yield highly accurate identification. We present data preprocessing, signature extraction, and matching techniques that are computationally inexpensive, and can scale to large datasets. We discuss implications of the attack and challenges associated with defending against such attacks.
\end{abstract}

\maketitle
\section{Background}
Studies of high-resolution images of the brain provide the basis for our understanding of the essential processes that underlie the functioning of the brain, help characterize behaviour and anatomy, and their dys-regulation due to disease, neurological disorders, and aging. Images of the brain, neuroimages, serve as phenotypes (observables) of structure and function. Various modalities of imaging rely on different biophysical processes and properties of the brain tissue. Consequently, these modalities highlight different aspects of the brain --  broadly speaking these images reveal anatomical structure or behavioral function. Anatomical images are high resolution (< 1 mm side) 3-D images that capture subtle structural details. Structural Magnetic Resonance Images (Structural MRI) can reveal clear boundaries between gray and white matter. Diffusion Tensor Imaging (DTI) and Diffusion Weighted Imaging (DWI) help visualize structural neuronal pathways. The physical manifestation of these features -- volumes of gray and white matter and direction of neuronal pathways,  are often dys-regulated in subjects with neurological disorders such as Attention Deficiency Disorder (ADD) and Attention Deficiency Hyperactive Disorder (ADHD), or neurodenerative diseases such as Alzheimer's and Parkinson's. 

Functional images are typically lower resolution 4-dimensional images (three spatial dimensions and one temporal dimension). Commonly used modalities for functional imaging include functional MRIs (fMRI), Electron Encephalography (EEG), and Magnetic Encephalography (MEG). Functional images capture dynamic behaviour of the brain, both in absence of stimulus (called resting-state) and in presence of stimulus (called task-specific). They reveal processes that underlie behaviour, provide the basis for learning, dys-regulation of typical features in subjects with neurological disorders, and aid in prediction (prognosis and diagnosis) of neurological disorders. EEGs use electrodes that record patterns of electrical activity in the brain. Anomalies in normal electrical activity are used to diagnose epilepsy \cite{Adeli07} and sleep disorders \cite{Petit04}. MEGs record magnetic fields resulting from electrical activity. They are used to understand perception, and in pre-surgical procedures to find regions of the brain associated with pathology of interest \cite{Anderson14}. Functional MRIs and MEGs have better spatial resolution than EEGs; conversely, EEGs have better temporal resolution.

In this paper, we focus on privacy aspects of functional MRI -- the most commonly used neuroimaging modality. Functional MRI measures local changes to quantify the activity of different regions of the brain -- for instance, Blood Oxygen Level Dependent fMRI (BOLD fMRI) measures the de-oxygenation of blood in veins to measure neuronal activation of surrounding regions. Functional MRIs are used both in research and in practice. They are used in pre-operative procedures to identify regions of the brain that are strongly correlated for various tasks. Similarly, ``cortical maps'' are constructed from fMRI images for planning surgeries. In research, fMRIs are used to study diverse phenotype, including Alzheimer's \cite{Clifford08}, schizophrenia \cite{Holmes05, Mitchell04}, bipolar disorder (BPD) \cite{Mitchell04}, depression \cite{Holmes05},dyslexia \cite{Ruff03}, hyperlexia \cite{Turkeltaub04}, homosexual pedophilia \cite{Schiffer08}, cocaine addiction \cite{Breiter97}, gambling \cite{VanEssen13}, minimal consciousness \cite{Moritz05}, extraversion, self-consciousness, romantic love \cite{Fisher05},  sexual arousal \cite{Schiffer08}, political leaning \cite{Westen06}, neurotism, extraversion and self-consciousness \cite{Eisenberger05}.

Structural and functional images from fMRIs are used to create connection maps of the brain, called connectomes (analogous to genome, which is a map of the gene). The study of connectomes is called connectomics (again, analogous to genomics).  A common, intuitive model for understanding brains is a correlation network, which represents regions of the brain that are co-activated in specific contexts. Connectomics provides insights into the architecture of these networks, and how the architecture (i.e., structure) is associated with different behaviour (i.e., function) \cite{Sporns05}. Connectomic studies are driven by, and drive both quantity and quality of neuroimages.

To summarize, large-scale repositories of high-resolution neuroimages are essential for critical clinical and behavioral studies. However, the highly sensitive nature of the data collected -- images of human brains is a significant cause for concern with respect to data privacy. Consequently, neuroimage studies anonymize subjects by removing all personally identifiable information. This simple step prevents a naive attacker. However, analogous to de-anonymization attacks in social networks, as we show in this study, neuroimages can be de-anonymized. De-anonmyziation of a subject in an external dataset can de-anonymize all other records of the same subject in other datasets. Such an attack would significantly compromise privacy of individuals. 

In this paper, we present a de-anonymization attack on neuroimages in the public domain. Our attack model assumes that there is an entity (the attacker) with a set of personally identifiable neuroimages (potentially from the past), and a public domain database of de-identified patient records that contains neuroimages. This is indeed the existing setup, where imaging centers have access to identifiable imaging data, and repositories from research efforts such as the Human Connectome Project release large amounts of de-identified neuroimaging data, metadata, and often, highly sensitive genomic data. We demonstrate that using our technique, imaging centers can reference imaging data across these datasets, and gain access to patient metadata and health records. We use methods based on matrix sampling and non-linear dimensionality reduction to identify subjects with high accuracy. Our proposed methods are shown to have provide theoretical guarantees, while working exceedingly well in practice, in terms of de-identification accuracy and computational cost. 

We make the following specific contributions in this paper:
\begin{itemize}
    \item Present the necessary mathematical framework for leverage score sampling and t-distribution Stochastic Neighbor Embedding (t-SNE) -- the core algorithms used to efficiently implement the proposed attack.
    \item Use both resting state and task-data in the HCP to infer the identity of individuals when one of the datasets has been de-anonymized. Specifically, we show that de-anonymization of a set of subjects performing one task can result in their identification, even if they are performing different tasks.
    \item Use anonymized task data in the HCP to identify the task being performed by a subject using t-distribution Stochastic Neighbor Embedding.
    \item Predict the efficacy with which an unknown subject is performing a given task from the de-identified fMRI.
    \item Show that the signatures can be used to de-anonymize subjects with neurological disorders such as Attention Deficit Hyperactive Disorder (ADHD).
    \item Simulate multi-site image acquisition and demonstrate the robustness of signature to different functional MRI machines.
\end{itemize}
Our demonstrated attack does not rely on any meta-data such as age, sex, ethnicity, education, intelligence, etc. Instead, it relies on patterns in the functional image, which are unique to  individuals. In view of our attack, we also highlight challenges associated with simultaneously defending against the attack, while retaining integrity of data in a way that supports necessary data analyses.
\section{Related Literature}
The topic of privacy and confidentiality is not new in neuroimaging studies. Data sharing between laboratories and open-data access for the public is essential for collaborative research, and for ensuring reproducibility. Large-scale efforts such as International Neuroimaging Data-sharing Initiative (INDI), Alzheimers Disease Neuroimaging Initiative (AFNI), and the Human Connectome Project (HCP) often need to strike a fine balance between enabling research, while respecting the privacy of subjects. A primary requirement of privacy is to prevent re-identification, or de-anonymization, which is one of the core objectives of the INCF Task Force on Neuroimaging Datasharing \citep{Poline12}.

In the United States, the Department of Health and Human Services (HHS) has specified standards for the privacy of identifiable health information. The legal requirements of these standards include removal of identifiable information, such as name, social security number (SSN), contact details, home address, etc. These standards are broadly drafted for all studies involving human subjects. The applicability of these standards to specific neuroimaging studies for medical doctors or non-medical scientists is complicated, and is described in detail by \citep{Tovino05}. Similar standards are established by organizations such as the National Institute of Health (NIH) and Organization for Economic Cooperation and Development (OECD) \citep{Poline12}.

The anonymization of subjects, as mandated by the HHS Privacy Rules, protects against naive attackers, but does not account for attacks by a competent adversary. Raw 3D images of the brain, regardless of modality -- structural MRI, CT scans, MEGs, etc. also include facial features. Each image can be thought of as an overlay of several 2D images, each of which has the outline of the subject's face, such as shape of nose, cheeks, eye sockets, along with the cranium that houses the brain. 
The problem of de-identification was first studied by \citep{Chen07}.  \citep{Budin08} showed that the chances of matching a 3D reconstructed face with the photograph of the subject was greater than what was expected by chance. With increasing resolution of strutural images, various techniques have been developed with the aim of overcoming threats to privacy. The process of skull stripping -- removing non-brain voxels, was developed well before this threat was first studied. Skull stripping classifies voxels as brain and non-brain, and masks the latter. This procedure is effective in removing the skull, when provided with a faceless image, but ineffective in presence of facial features, since the algorithms confuse facial voxels and brain voxels. \citep{Bischoff-Grete07} describe a procedure for de-identifying images by removing voxels that correspond to pre-defined facial features. \citep{Budin08} propose deformation of facial features to obfuscate the face, however the effect of this transformation on the integrity of brain voxels is unknown. \citep{Milchenko13} describe a more conservative approach that only deforms the surface of the face. Another commonly used strategy is to blur the face, instead of removing the voxels altogether \citep{Avram15}. However, all of these procedures have been shown to have an adverse affect on the quality of images, with respect to downstream analyses.

The security and privacy implications of functional MRI data are relatively unexplored in the security community, particularly as they relate to existence, identification, and use of brain signatures. Brain signatures or brain fingerprints are patterns in the functional brain image that are unique to  individuals. \citep{Finn15} show that two functional connectomes belonging to the same subject are more similar than functional connectomes drawn from two different subjects. It has also been shown that these distinct profiles can be used to predict fluid intelligence. These findings are important to precision psychiatry, a discipline whose goal is to find key individual differences, which in turn aids in the generation of models that can predict behavior.  However, there are serious privacy concerns, since the accuracy in identification was shown to be around 90\%, and when restricting the analysis to the parieto-frontal region, the accuracy of identification is close to 100\%. Similarly, \citep{Vanderwal17} show that individual differences in functional connectivity when viewing fast-paced movies can be used to identify individuals with close to 100\% accuracy. A leverage-score based sampling method was first presented by \citep{Ravindra18}. Leverage-score sampling identified a small set of discriminative features that strongly expressed the signature. These features are used to identify small regions, called parcels on the cerebral cortex, which are important to the task of identification. 

In this paper, we use leverage score sampling, along with an alternate technique based on t-distribution stochastic neighbor embedding for various identification tasks. Our methods offer a number of advantages over state of the art techniques: first, they offer higher accuracy in identification; second, our method automatically selects features from connectomes, without requiring any knowledge of underlying electro-physiological processes or anatomical structures; third, the features selected by our method are shown to be robust across populations of subjects. For these reasons, our methods provide an excellent basis for a de-anonymization attack.

\begin{figure}[h]
    \centering
    \includegraphics[scale=0.6]{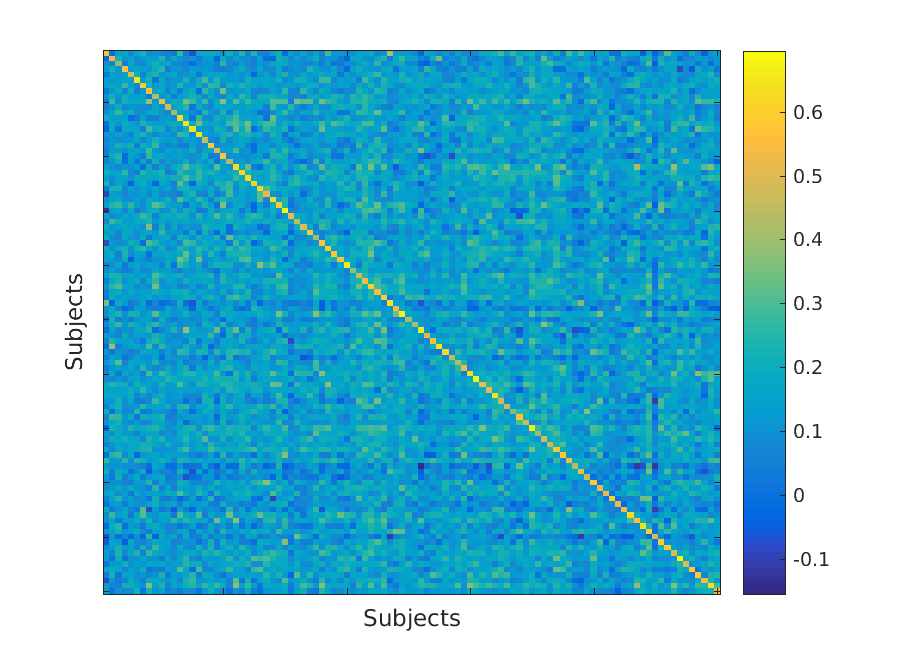}
    \caption{Pairwise similarity of resting-state connectomes. The diagonal entry in this matrix represents the similarity between two sessions of the same subject. All off-diagonals represent similarity of connectomes between two different subjects. The high values on the diagonal show that connectomes of the same subject are more similar than connectomes belonging to different subjects.}
    \label{fig:rest_v_rest}
\end{figure}

\begin{figure}[h]
    \centering
    \includegraphics[scale=0.6]{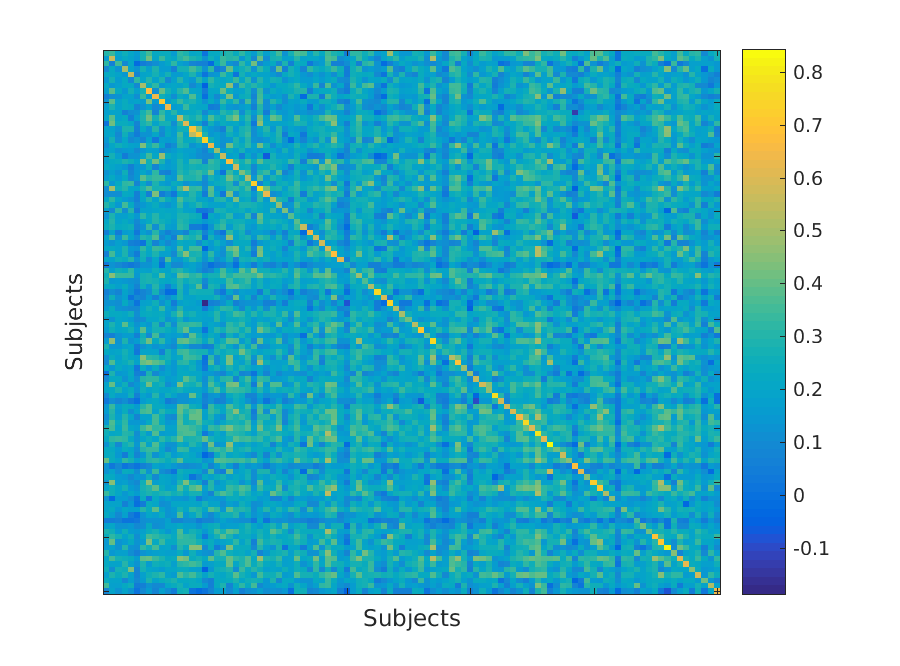}
    \caption{Pairwise similarity of task connectomes for language processing task in the HCP. The high diagonal values denote strong similarity between two connectomes belonging to the same subject. However, the contrast between diagonal and off-diagonal values is weaker than in resting-state.}
    \label{fig:lang_v_lang}
\end{figure}

To clearly characterize the privacy threat, we compute the similarity matrix obtained by applying the method of Ravindra et al. \citep{Ravindra18} on the resting-state functional MRIs from the Human Connectome Project (HCP) (Section \ref{sec:signature}). The results are visualized in Figure \ref{fig:rest_v_rest}. The strong intra-subject similarity is represented by the high values on the diagonal; conversely, the weak inter-subject similarity is represented by low off-diagonal values. This pattern holds even in case of task functional MRI, though the contrast between diagonal and off-diagonal values is less stark, as shown in Figure \ref{fig:lang_v_lang}.

These results suggest that if one publicly available dataset is de-anonymized, or if a patient's record at one hospital is de-anonymized, then we can accurately de-anonymize any other corresponding patient datasets or patient records. This potentially reveals confidential details such as the presence of, or progression of disease, behavioural traits, and access to corresponding non-imaging data, such as age, sex, contact details, gender identity, sexual orientation, etc. This issue is of greater concern, especially for multi-modal datasets such as the HCP and Enhancing Neuro Imaging Genetics through Meta Analysis (ENIGMA), which also share Genome-Wide Association Studies (GWAS) data. GWAS is used to find genetic variants, and to predict genetic predisposition to diseases. This is an increasingly important threat, since neuroimage datasets are increasingly enriched by patient details that are relevant to clinical and research studies.

\nocite{Waller17}
\section{Brain Signatures}
\label{sec:signature}

This section is organized as follows: first, we describe various mathematical models and methods that form the basis for our de-anonymization attack. We then describe the Human Connectome Project (HCP) and ADHD-200 datasets on which we demonstrate our attack. We also describe the role of the canonical Human Brain Atlas in this process. Finally, we perform a comprehensive experimental study on all aspects of our attack, its accuracy, and scope.

\subsection{Methodology}

We assume an attacker has two functional MRI datasets in their possession and that one of them is de-anonymized. The attacker wishes to de-anonymize the other dataset. However, such an attacker would first need to convert raw neuroimages into a format suitable for analysis, and then extract individual-specific brain signatures from these processed images. 
%
A functional MRI is a 4D image. It is comprised of 3 spatial $(x,y,z)$ dimensions, and one temporal dimension $(t)$.  For a given time point, an fMRI is a 3D snapshot of the brain. Each 3D unit is a voxel (a volume pixel). Similarly, for a given spatial location, fMRI measures the activity of the corresponding voxel across time. This signal corresponds to a ``time-series''. 

A raw functional MRI has spatial and temporal artifacts. We apply a denoising procedure that corrects for head motion by the subject during image acquisition, removes skull voxels, and corrects for non-homogeneous magnetic fields. We account for difference in sizes of brains by registering all brains to a standard brain. This preprocessing procedure is described in detail in Section \ref{subsection_preprocessing}. The output of this step is a denoised, normalized image. We group voxels in this image into regions (or parcels) using a standard brain atlas (Section \ref{subsection_atlas}). We then compute the average time-series for each region by averaging over all voxels.
We organize this data into a 2D matrix, in which each row corresponds to the averaged time-series associated with a region ( the matrix has dimensions {\it regions $\times$ time-points}). We quantify the co-activity between different regions using correlation measures to create a {\it regions $\times$ regions} similarity matrix, which is a representation of the functional connectome. The procedure is repeated for all images in the database.

Each element in the similarity matrices can be viewed as a feature of the corresponding connectomes, and the goal is to find a compact set of features that distinguish subjects from each other. To accomplish this, we vectorize the similarity matrix by stacking its columns one on top of the other. In this manner, each 4D image is reduced to a single (column) feature vector with $regions^2$ elements. We take all of the feature vectors corresponding to images in our de-anonymized set and organize them into a matrix. Each column of this matrix corresponds to a subject, and each row corresponds to a feature from the similarity matrix. We compute a second matrix in a similar manner from all of the images in the anonymous image dataset. The goal is to use these matrices to de-anonymize the second dataset.
We now introduce two techniques for carrying out such an attack.


Our first technique is based on leverage-score sampling. At a high level, leverage-scores of  a matrix measure relative importance of rows/ columns. Recall in our setting that a row of our group matrix corresponds to individual entries (features) in the similarity matrix. We use leverage scores to identify important components (or subspaces) in these vectors, and only retain the rows (features) with high scores. The output of this procedure is a reduced group matrix with (far) fewer rows; however these rows are important for discriminating between individuals. Finally, we measure similarity between all pairs of individuals across the two subject groups using the reduced group matrices based on Pearson Correlation. Pairs of subjects with high correlation correspond to predicted matches. In section \ref{subsection_sampling}, we formally define leverage-scores and the algorithm for leverage-score sampling. This technique is used to de-anonymize healthy subjects at rest and while performing tasks. We also demonstrate their effectiveness of this technique in de-anonymizing patients with ADHD.

Our second algorithm is based on t-distribution stochastic neighbour embedding (t-SNE). This non-linear dimensionality reduction technique is used for clustering and visualization. We use t-SNE on the previously described group matrices to reduce the number of rows down to two. The clusters obtained thereof are used to separate various tasks and also predict the nature of task from a new, anonymized image. The details of our t-SNE based method are presented in Section \ref{sec:tsne}. We show that t-SNE is particularly effective for our application because it maintains pairwise distance in low dimensions well, while maintaining underlying cluster structure.


\subsubsection{Models and Methods}
\label{subsection_models_methods}
\begin{figure*}
    \centering
    \includegraphics[width=\textwidth]{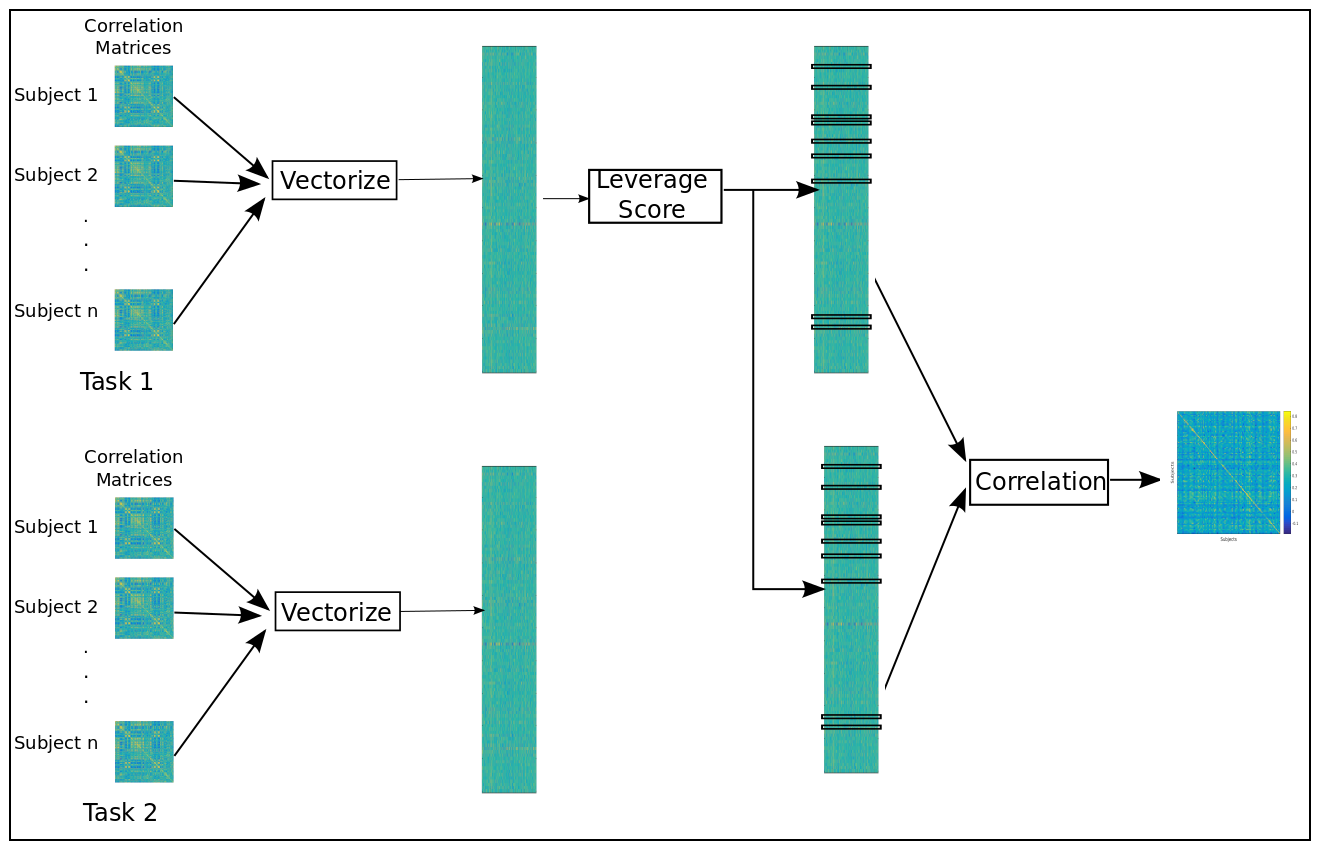}
    \caption{\textbf{Description of the workflow and the various matrices}. The time-series correlation matrices of each subject are vectorized to create the group matrices. The feature-space is restricted to the top leverage scores of the first group matrix. The features corresponding to the top leverage scores are highlighted. The inner-product (correlation) between pairs of subjects in this reduced feature-space is the basis for classification.}
    \label{fig:my_pipeline}
\end{figure*}

\paragraph{Generating Functional Connectomes.} 
Given raw fMRI images, the general pre-processing pipeline, shown in Figure \ref{fig:fmri_preprocessing_pipeline}, removes temporal and spatial artifacts from raw images. These de-noised images are collections of time-series data, with each time series having an associated position in 3-dimensional space. 
The time-series matrix ({\it voxel $\times$ time}) is z-score normalized, averaged over all voxels in a region (as defined by an atlas), and used to compute correlations (using Pearson correlation) to generate a dense co-firing matrix. Note that this matrix can also be interpreted as a weighted complete graph, where nodes correspond to regions and edge weights correspond to correlation in neuronal activity.


A variant of this preprocessing pipeline is used in most fMRI studies.
To demonstrate a de-anonymization attack, we require two datasets -- a dataset that is de-anonymized, and the second target dataset. We will use the identities revealed for connectomes in the first dataset to predict the corresponding identities in the target dataset. In our experiments on the HCP dataset, we use the left-to-right encoding (L-R) of a set of 100 subjects to create the first dataset; we use the right-to-left enconding (R-L) of the same subjects to create the target/second dataset. The encodings arise from different ordering of scans, which we describe in Section \ref{subsection_dataset}. We vectorize each correlation matrix and stack them into two matrices, one per dataset as shown in Figure \ref{fig:my_pipeline}. We note that the symmetry of the correlation matrix allows us to vectorize only the top (or bottom) triangle of the correlation matrices. 

We use matrix sampling to identify features that encode distinguishing signals. By restricting the predictive procedure to these features, we obtain higher performance in identification, while also making prediction faster. In resting-state fMRI dataset in HCP, this procedure is used to reduce the feature-space from $64620$ to $< 100$ rows. In our second set of experiments, we use a non-linear dimensionality reduction technique (t-SNE) to distinguish between tasks; however the massive reduction in dimension results in loss of individual specific signals and hence cannot be used in specific applications where the leverage-score based method is found to be superior. To this end, our two proposed methods are complementary.

\subsubsection{Row sampling for identifying location of signatures}
\label{subsection_sampling}

In our description of the experimental setup in Section \ref{subsection_models_methods}, we introduce group-level matrices comprised of vectorized functional connectomes. Let us call this matrix $A$. In our example, $A$ has $64620$ rows (derived from 360 regions of the brain from the atlas for $(360 \times 359)/ 2 = 64620$ region-pair correlations or features), with each row representing one entry from the region-wise correlation matrix. $A$ has 100 columns, one for each subject. 

Given a matrix $A \in \mathbb{R}^{m \times n}$, the problem of row-sampling is to choose a subset $s$ of rows, $s < m$ such that the resulting matrix $\tilde{A} \in \mathbb{R}^{s \times n}$ approximates $A$. Since $A$ and $\tilde{A}$ are of different dimensions, the error in approximation is expressed as $||A^T A - \tilde{A}^T\tilde{A}||$. 
Row sampling can be viewed as a method for dimensionality reduction. However, it is fundamentally different from other well-known approaches such as Principal Component Analysis (PCA). PCA uses the eigenvectors of the covariance matrix as basis vectors. However, eigenvectors themselves have little physical interpretability, since the original data vector has no interpretable representation in the eigenvectors. On the other hand, row-sampling techniques allow us to retain original data vectors, and can therefore be used for feature selection. Our goal is to find a small subset of rows that strongly express the signal required to discriminate subjects (i.e., the columns).

A randomized meta-algorithm that takes a matrix $A \in \mathbb{R}^{m \times n}$ and outputs a sketch matrix $\tilde{A}$ is presented in Algorithm \ref{alg:meta_alg}~\citep{Drineas06}.

\begin{algorithm}
\begin{algorithmic}[1]
\Function {row\_sample} {A,s}
\State {Let $\tilde{A}$ be an empty matrix}
\For {$t = 1$ to $s$}
\State {Randomly sample a row according to the distribution $P$}
\State {Let $A_{i_t,\star}$ be the sampled row, with corresponding probability $p_i$}
\State {Set $\tilde{A}_{t,\star} = \frac{1}{\sqrt{s p_i} }A_{i_t,\star}$}
\EndFor\\
\Return $\tilde{A}$
\EndFunction
\end{algorithmic}
\caption{Meta-algorithm for row-sampling}
\label{alg:meta_alg}
\end{algorithm}

In this algorithm, each row of the matrix is associated with sampling probability $P$s. In each of $s$ iterations, a row of $A$ is picked in i.i.d (independently and identically distributed) trials. The re-scaling of $\tilde{A}$ in line 6 ensures that $\tilde{A}^T \tilde{A}$ is unbiased, i.e., $\mathbb{E}[(\tilde{A}^TA)_{i,j}] = (A^T A)_{i,j}, \forall i \in \{ 1 \dots m\}, j \in \{ 1 \dots n\}$ \citep{Drineas06}

The theoretical guarantees and practical performance of this meta-algorithm hinge on a cleverly chosen distribution $P$. For instance, a Uniform Random distribution performs poorly in practice. On the other hand, a probability distribution biased on norms of the rows has an intuitive appeal, since important features would be associated with higher weights. This leads to $l_2$ sampling, wherein $P$ is defined by the 2-norm of rows.

\begin{equation}
p_i = \frac{|| A_{i,\star} ||_2^2}{\sum_i || A_{i,\star} ||_2^2} = \frac{|| A_{i,\star} ||_2^2}{|| A_{i,\star} ||_F}.
\end{equation}
If the sampling is weighted by $l_2$ norm and we select $\mathcal{O} (m\ \log{m})$ rows, Drineas et al.~\citep{Drineas06}  show that: 
\begin{equation}
\mathbb{E}[|| A^T A - \tilde{A}^T\tilde{A}||_F] \leq \frac{1}{\sqrt{s}}||A||_F^2.
\label{two_norm_ineq}
\end{equation}
This suggests that $\tilde{A}$ can be used as a low-rank approximation to $A$. However, the error bound defined herein is additive. A stronger error-bound is one in which the error is relative, i.e., the approximation is within a factor $(1 \pm \epsilon)$ of the true value. A compact yet powerful method to achieve this is based on a concept of leverage scores.

\paragraph{Leverage Scores.}
Given a matrix $A \in \mathbb{R}^{m \times n}$ with $m \gg n$, let $U \in \mathbb{R}^{m \times n}$ be an orthogonal matrix that spans the column space of $A$. For convenience, we assume that $A$ has full column-rank. $U$ has two properties:
\begin{enumerate}
    \item{$U^T U = I$ (as $U$ is orthonormal)}
    \item{$U U^T = P_A$ (a projection matrix that spans the column-space of $A$)}
\end{enumerate}
In manner similar to Equation \ref{two_norm_ineq}, we can define a probability distribution on the rows of $U$. These probabilities are known as \textit{leverage scores}.
\begin{equation}
p_i = \frac{|| U_{i,\star} ||_2^2}{\sum_i || U_{i,\star} ||_2^2}= \frac{1}{n}(P_A)_{i,i} \qquad \forall i \in \{1 \dots m\}.
\label{leverage_probability_eqn}
\end{equation}

If we select $\mathcal{O}(k \log{k}/ \epsilon^2)$ rows, we get a relative error bound as follows:
\begin{equation}
    || A - A \tilde{A}^\dagger \tilde{A}||_\zeta^2 \leq (1+\epsilon) ||A - A_k||_\zeta^2,
\end{equation}
where $\zeta \in \{2,F\}$, $\epsilon \in [0,1/2)$, $A_k$ is the best rank-k approximation and $\dagger$ represents the pseudo-inverse (\citep{Drineas08}). These randomized algorithms provide the theoretical basis for our de-anonymization approach, which samples the top-leverage scores in a deterministic manner, which provides good matrix sketches \citep{Cohen15, Papailiopoulos14}. In Ravindra et al. \citep{Ravindra18}, this method is called \textit{Principal Features Subspace} Method.

\paragraph{Principal Features Subspace Method}
\label{subsection_principal_features_subspace}
As before, let $A$ be the matrix of connectomes, and $U$ be the orthonormal matrix that spans the column space of $A$. Additionally, let $t$ be the number of features that need to be retained. An example of such a matrix $U$ is constructed using left singular vectors from a Singular Value Decomposition (SVD) of $A$. We can then compute the leverage($l$) scores of $A$ as:
\begin{equation}
l_i = || U_{i,\star} ||_2^2, \qquad \forall i \in \{1 \dots m\}.
\end{equation}
We sort the leverage scores and retain the features corresponding to the top $t$ leverage scores. We call this subspace the \textit{principal features subspace}. In contrast to prior randomized approaches, we select features in a deterministic manner; Cohen et al.~\citep{Cohen15} provide theoretical bounds for this selection process.

Starting from the matrix of vectorized correlation values $A$, we compute the left singular vectors using SVD. The ordering of edges according to their leverage scores, if robust across different groups, is indicative of a set of features  that can accurately fingerprint an individual's functional connectome. In this case, for a given parcellation scheme, and for a given measure of region-to-region coherence, we need to apply SVD just once to determine the relevant edges. 

\subsubsection{t-Distribution Stochastic Neighbours Embedding (t-SNE)}
\label{sec:tsne}

The goal of t-SNE is to map high-dimension data-points into a low-dimension space, while maintaining the cluster structure of the data. It relies on the assumption that high-dimensional data lies on several
different, but related, low-dimensional manifolds. For instance, images of objects from multiple classes
seen from different viewpoints. t-SNE is commonly used in visualization, since it can be used to map data down to a visual plane (i.e., in two-dimensions). We now describe the use of t-SNE for de-anonymization.

Stochastic Neighbours Embedding (SNE) converts pairwise distances between points in a dataset to conditional probabilities that represent their degree of similarity. For any two high-dimensional data points $x_i$ and $x_j$,

\[  p_{j|i} =  \left\{
\begin{array}{ll}
      \frac{\exp(-|| x_i - x_j ||/2\sigma_i^2)}{\sum_{i \neq k}\exp(-|| x_i - x_k ||/2\sigma_i^2)} & x_i \neq x_j \\
      0 & otherwise\\
\end{array} 
\right. \]

Here, $\sigma_i$ is the variance of a Gaussian centered at $x_i$. Equations of the same form can be written for pairs of points $y_i$ and $y_j$ in the low-dimension embedding. We denote the associated conditional probability as $q_{j|i}$. A mapping that respects cluster structure needs to map pairwise distances of points. Hence, a good mapping would result in the same values for $p_{j|i}$ and $q_{j|i}$. An intuitive measure, which would penalise any deviation of $q$ from $p$, is the Kulbeck-Leiber divergence, which is defined as:
\begin{equation}
    C = KL(P_i||Q_i) = \sum_j p_{j|i} \log \frac{p_{j|i}}{q_{j|i}}
    \label{eqn:sne_cost_function}
\end{equation}

The only unspecified detail here is the procedure to set $\sigma_i$. The variance about a point in a sparse neighborhood is much larger than the variance about a point in a dense neighborhood. Any choice of $\sigma_i$ induces a probability distribution $P_i$ over all other data-points. Hence, control on the entropy of $P_i$ can be exercised by controlling $\sigma_i$. SNE models $P_i$ on the basis of a user-defined quantity, perplexity:
\begin{equation}
    Perp(P_i) = 2^{H(P_i)}
    \label{eqn:perp}
\end{equation}
where, $H$ is the entropy defined as:
\begin{equation}
    H(P_i) = - \sum_j p_{j|i} \log_2 p_{j|i}
    \label{eqn:entropy}
\end{equation}

SNE minimizes the KL-divergence using a gradient descent algorithm. The gradient descent algorithm is remarkably simple. Differentiating the cost function in \ref{eqn:sne_cost_function}, we get: 
\begin{equation}
    \frac{\partial C}{\partial y_i} = 2 \sum_j (p_{j|i} - q_{j|i} + p_{i|j} - q_{i|j}) (y_i - y_j)
\end{equation}
The $(y_i - y_j)$ term in the gradient can be interpreted as the vector along which a force is felt between the pair of points in high dimension. The sign of the coefficient represents whether the force applied is attractive or repulsive. If the high-dimensional points and their corresponding low-dimension map are similar, the magnitude of the force felt is negligible. 

SNE is conceptually simple, however, it has several shortcomings. First, the KL divergence is assymmetric, i.e., $KL(P||Q) \neq KL(Q||P)$, making it computationally expensive. Second, SNE suffers from the problem of ``data-crowding'', which we will explain with an example: if the data is inherently in a certain space (say in 10 dimensions), and SNE is used to map the data down to two dimensions, then this map may not be not faithful. For instance, 11 datapoints can be equidistant in 10 dimensions, but geometry forbids this in two dimensions (a maximum of 3 points can be equidistant in 2 dimensions). Beyond data crowding, the variance $\sigma_i$ needs to be estimated for each data point. 

The t-SNE method, first introduced by van Der Maaten \citep{vanDerMaaten08} is designed to overcome these problems. It overcomes the assymetry problem by minimizing a single KL-divergence as follows:
\begin{equation}
    C = KL(P||Q) = \sum_i\sum_j p_{ij} \log \frac{p_{ij}}{q_{ij}}
    \label{eqn:tsne_cost_function}
\end{equation}
This is symmetric because the joint probability distributions are symmetric, i.e., $p_{ij} = p_{ji}$ and $q_{ij} = q_{ji}$, $\forall i,j$. The probability distributions themselves are given by:
\[p_{ij} =   \left\{
\begin{array}{ll}
       \frac{\exp(-|| x_i - x_j ||/2\sigma^2)}{\sum_{k \neq l}\exp(-|| x_k - x_l ||/2\sigma^2}) & x_i \neq x_j \\
      0 & otherwise
\end{array} 
\right. \]
This definition of $p_{ij}$ is potentially problematic for outliers. If $x_i$ is an outlier in high-dimension, then $p_{ij}$ is very low for all other points $j$. Hence, the position of the point in the low-dimension map does not affect the cost function. Hence, its position is not well determined with respect to positions of other points in the dataset. This is avoided by setting $p_{ij} = (p_{i|j} + p_{j|i})/2$, thereby ensuring that $\sum_jp_{ij} > 1/2n, \forall j$. This implies that each point exerts some control over the cost function, making it robust to outliers.   

We can also define $q_{ij}$ in a similar fashion. Here, the variance is estimated once for the entire dataset, hence we circumvent having to estimate the variance of a Gaussian about each point. To overcome the data-crowding problem, t-SNE uses the student t-distribution with one degree (also called the Cauchy distribution) to define the probability distribution in the low-dimension space. This distribution differs from a Gaussian in that it has a heavier tail. Hence, pairs of points at moderate distances in the high-dimensional space can be faithfully mapped to low-dimension, thereby preventing unwanted attractive forces between such points. Formally: 
\begin{equation}
    q_{ij}  = \frac{(1+|| y_i - y_j||^2)^{-1}}{\sum_{k \neq l}(1 + ||y_k - y_l||^2)^{-1}}
    \label{eqn:tsne_qij}
\end{equation}
This equation has a desirable form because $(1+||y_i - y_j||^2$ approaches the inverse-square law for large $||y_i - y_j||$, hence it is robust to changes in the scale of the map for map-points that are far away from each other. The gradient for t-SNE is slightly different, since we use both Gaussian and t-distribution.
\begin{equation}
    \frac{\partial C}{\partial y_i} = 4\sum_{j} (p_{ij} - q_{ij})(y_i - y_j)(1+ ||y_i - y_j||^2)^{-1}
    \label{eqn:tsne_gradient}
\end{equation}
We can now assemble all the pieces into a simplified version of the t-SNE algorithm. 
\begin{algorithm}
\begin{flushleft}
\textbf{Input:} $X = \{x_1, x_2, \dots , x_n \}$\\
\textbf{Input:} Perplexity $Perp$,\\
\textbf{Input:} Number of iterations $T$,\newline
\textbf{Input:} Learning rate $\eta$, Momentum $\alpha(t)$\newline
\textbf{Output:} $Y = \{y_1, y_2, \dots , y_n \}$\\
\end{flushleft}
\begin{algorithmic}[1]
\State {Compute $p_{j|i}$ from perplexity, as given in equations \ref{eqn:perp} and \ref{eqn:entropy}}
\State {$p_{ij} = \frac{p_{i|j} + p_{j|i}}{2}$}
\State {Initialize $Y^{(0)}$ from $\mathcal{N}(0, 10^{-4}I)$}
\For{t = 1 to T}
    \State {Compute $q_{ij}$ from Equation \ref{eqn:tsne_qij}}
    \State {Compute gradient as per Equation \ref{eqn:tsne_gradient}}
    \State {$Y^{(t)} = Y^{(t-1)} + \eta \frac{\partial C}{\partial Y} + \alpha(t)  (Y^{(t-1)} - Y^{(t-2)})$}
\EndFor
\end{algorithmic}

\caption{t-distribution Stochastic Neighbour Embedding}
\label{alg:tsne}
\end{algorithm}

\subsection{Dataset}
\label{subsection_dataset}

The Human Connectome Project (HCP) Consortium provides a large collection of neuroimage data over the public domain. The Washington University-University of Minnesota (WU-Minn) project, which is part of HCP conducted a Healthy Young Adult study \citep{VanEssen13} resulting in 3T structural and functional MRI for 1113 adults, 7T resting and task magnetoencephalography (MEG) from 184 subjects, and 3T and 7T diffusion data. In this paper, use resting state functional MRI and task functional MRI for seven tasks.

The resting state functional MRI is acquired over the course of two sessions, spread across two days. The duration of each session was 30 minutes, with the first half session corresponding to L-R encoding and the second half session corresponding to R-L encoding. The voxels are isotropic, with side $2 mm$; the temporal resolution (TR) is 720 ms.  In this paper, we use data from 100 ``unrelated subjects'' to avoid genetic factors from confounding the results. For a detailed description of the acquisition protocol, see Smith et al. \citep{Smith13}.

The HCP image acquisition procedure collected task based functional MRI in addition to resting state functional MRI. The first session consisted of one session of rest, followed by working memory, gambling, and motor tasks. The second session consisted of another session of resting state activity, followed by language skills, social cognition, relational processing, and emotional processing. The tasks themselves comprised of visual and auditory stimuli presented to the subject, who could respond by means of buttons placed near their hands. The responses were recorded, and meta-data such as measures of performance were generated. We refer interested readers to Barch et al. \citep{Barch13} for justification of choices of experiments and details of each experiment.

The ADHD-200 dataset was collected by the International Neuroimaging Datasharing Initiative (INDI), as part of a collaboration of eight imaging sites around the world. It consists of multiple scans from 362 children and adolescents diagnosed with ADHD, along with 585 controls. The publicly available dataset consists of both structural MRI, resting state functional MRI, and associated phenotypic information \cite{Pierre17}. 

\subsubsection{Preprocessing}
\label{subsection_preprocessing}

\begin{figure*}
    \centering
    \includegraphics[width=\textwidth]{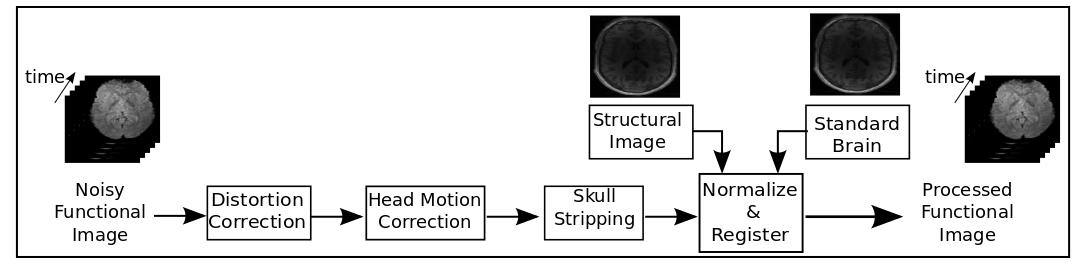}
    \caption{{\bf A typical functional MRI pre-processing pipeline.} This figure shows the general pre-processing steps that are performed on functional MRI, in order to correct for spatial and temporal artifacts. The exact order of the processing steps may sometimes change; sometimes extra steps such as slice-time correction may be added, depending on the quality of images and the acquisition protocols.}
    \label{fig:fmri_preprocessing_pipeline}
\end{figure*}

As with all images, pre-processing helps to clean the data of spurious patterns that arise due to technical, biological, and acquisition protocols.  In this paper, we follow the ``minimum pre-processing pipeline'' prescribed by Glasser et al. \citep{Glasser13} and Smith et al. \citep{Smith13}. The generic steps are shown in Figure \ref{fig:fmri_preprocessing_pipeline}.

Spatial pre-processing removes artifacts that affect the comparison of a given voxel across time. This includes correction for spatial distortions due to gradient non-linearity, which occurs due to local non-uniformity of the magnetic field. This is followed by head motion correction, which is vital because subjects invariably move their head during the course of imagine. Then, the functional MRI is registered to the subject's own structural MRI, to ascertain the structural membership of voxels in the functional image. To facilitate comparison across subjects, factors such as differing sizes of brains, difference in sizes of regions in brains need to taken into account. This is achieved by mapping all structural MRIs to a standard brain model, called the Montreal Neurological Institute (MNI) space. Then, voxels in the surface of the brain, called the cortex are separated from the sub-cortical volumetric voxels into a standard CIFTI format. The rationale behind storing data in this format is explained in Glasser et al.~\citep{Glasser13}. In this paper, we only use the cortex for identification. 

The HCP temporal pre-processing pipeline suggests a minimal high pass filter (cutoff of 2000 seconds) and a slow roll off beyond that, so as to achieve de-trending of data. In task MRI, the cutoff was 200 seconds. Temporal pre-processing is extremely important in studies that rely on correlation between time-series, which is the case in our work. In resting state functional MRI, the signals of interest correspond to fluctuations of low frequency that arise due to haemodynamic responses to neural activation. Hence, we use a bandpass filter between 0.008 Hz and 0.1 Hz. We also apply global signal regression on resting state data. This procedure removes signal-components that are expressed uniformly throughout the brain.

For the ADHD-200 dataset, we use the Burner pipeline, as outlined by Ashburner et al. \cite{Ashburner13}. Broadly speaking, the steps followed are in line with Figure \ref{fig:fmri_preprocessing_pipeline}. We refer interested readers to \url{http://www.preprocessed-connectomes-project.org} to download the preprocessed files from different pipelines.

\subsubsection{Brain Atlas}
\label{subsection_atlas}
A brain atlas is an annotated standard-brain image. Each brain voxel in an atlas is associated with a label. Atlases are necessary in fMRI analyses due to the unreliability of single voxel time-series, which is due to a combination of factors. First, the process of registration and the other pre-processing steps may alter the values of a time series. Second, voxels are created without regard to anatomical or electro-physiological properties of the brain. Ideally, each voxel is given exactly one label. This would make the atlas non-overlapping. Another desirable property is that labels are localized to small, functionally similar regions -- called parcels. The process of generating labels can either be done manually, or can be automated. Manual labels are high quality, as the annotations are provided by experts; however, they are slow and expensive. Atlases can also be automatically generated. For instance, a simple scheme to create an atlas with 1000 ROIs (regions of interest) would involve sampling 1000 voxels from a uniform distribution. Then, the regions are progressively grown on the basis of functional similarity and/or physical proximity.

Given a connectome, expressed as a $voxel \times time$ matrix and a membership function for each voxel, one can collapse it into a $region \times time$ matrix, simply by computing region-wise average of time series data. These parcels or regions also serve as a useful abstraction, since they allow us to think of the brain as having distinct regions. In this paper, we use the atlas due to Glasser et al. \citep{Glasser16}, since the atlas itself was developed on the the same HCP dataset. Additionally, it has been shown that this atlas is sufficient to capture patterns that are unique to individuals (Ravindra et al. \citep{Ravindra18}). The atlas partitions the brain into 360 regions, symmetrically distributed across the left and right hemispheres. Similarly, we use the AAL2 atlas due to Tzourio-Mazoyer et al. \cite{Mazoyer02} for the ADHD-200 dataset. The use of two different atlases on two different datasets shows the robustness of our methods to different atlases and datasets.

\subsection{Experimental Evaluation}

In this section, we demonstrate de-anonymization attacks under different scenarios using our methods. First, we show that de-anonymization of a subject performing one task can make them vulnerable to de-anonymization in any other dataset, where they may be performing different tasks. We also show that we can predict the nature of task being performed by the subject. Furthermore, we show that the signatures obtained for a task can be used to predict their efficacy in performing the task. These results are important in demonstrating the extent to which de-anonymization can compromise the privacy of healthy individuals who participate in neuroimaging studies. However, the more serious problem is with hospital records, which typically have subjects with neurological disorders. In this context, we show that we can de-anonymize subjects with neurological disorders such as ADHD. Finally, we simulate the most realistic scenario, wherein patients have images obtained from different MRI machines. Even in this case, we show that we can identify subjects with high accuracy.

\subsubsection{De-anonymization of subjects across tasks}

 We first demonstrate an efficient de-anonymization attack in task functional MRI. We assume that one data-set is de-anonymized, and we are interested in characterizing its impact on the anonymity on other datasets. The tasks performed by subjects in the two datasets are potentially different. For instance, the REST1 may be de-anonymized and we are interested in finding the correspondences in the LANGUAGE task. The accuracy in de-anonymizing a dataset of resting-state functional MRIs in the HCP was found to be in excess of 94\%, as shown in Figure \ref{fig:rest_v_rest}. However, task fMRIs are typically more complex, since task driven brain activities are more complex than spontaneous firings recorded in resting state sessions. Furthermore, task-based activations are localized to specific regions and lobes that are responsible for performing the task. For instance, a task with only visual stimuli would activate the visual cortex, but not the auditory cortex. 

As described in Section \ref{subsection_models_methods}, we construct two group-wise matrices. The first group has the L-R encoding of REST1, and the L-R encoding of the seven tasks. In the second group, we  have the R-L encoding of REST2, along with the R-L encoding of the same seven tasks. We stack the vectorized network representations of the images into two matrices -- one for each group. We assume that each of the datasets in the first group is de-anonymized and our task is to reveal the corresponding images in the second group. The feature space of both groups are restricted to the features with top leverage scores in the first group. Figure \ref{fig:my_pipeline} shows a succinct representation of this process. 

\begin{figure}[h]
    \includegraphics[scale=0.6]{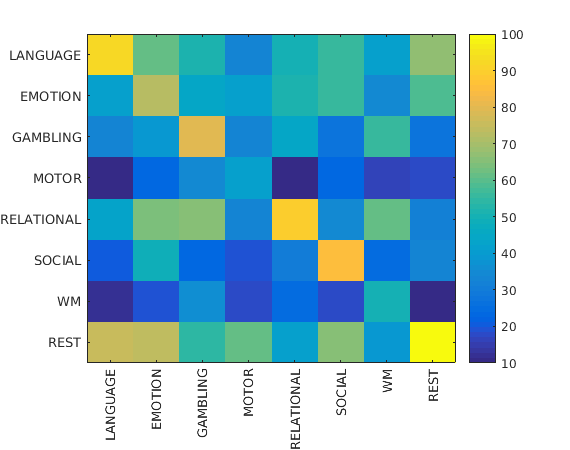}
    \caption{Identifiability of subjects performing different tasks. The rows consist of de-anonymized datasets and columns consist of datasets in which subjects are anonymous. This figure shows the effect of de-anonymizing different datasets.}
    \label{fig:identifiability_across_tasks}
\end{figure}

The results for this experiment can be seen in Figure \ref{fig:identifiability_across_tasks}. It shows that the accuracy of identification is stronger along the diagonal, i.e., when both the datasets correspond to images of the same task. The accuracy is particularly high (> 94\%) when they both are resting state functional images. Similarly, accuracies for language processing and relational processing tasks are $> 90\%$, and for social processing task is $> 80\%$. Surprisingly, MOTOR and WM tasks are ineffective in predicting the correspondence, even for the same task. We also can see that the matrix is clearly asymmetric, implying that de-anonymization of different tasks has varying impact on the anonymity of other datasets. 
On the whole, we observe that if a resting-state dataset is de-anonymized, it can be used to de-anonymize other tasks with high accuracy. This is particularly worrisome, since most studies produce resting state functional images. On the other hand, de-anonymizing tasks such as working memory and motor tasks reveals relatively less about the identities of individuals in other datasets. 


\subsubsection{Prediction of task}

\begin{figure}
    \centering
    \includegraphics[width=\columnwidth]{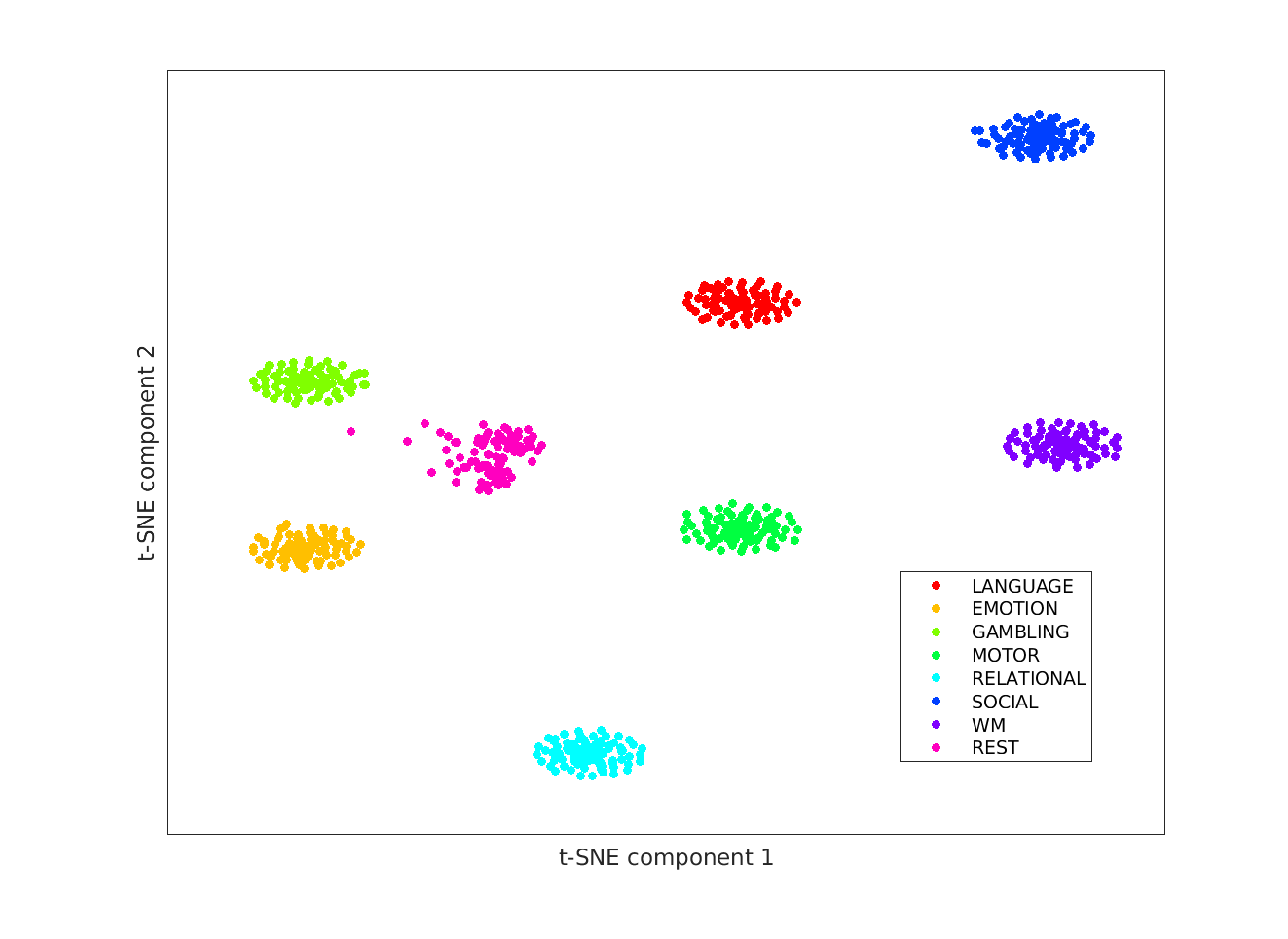}
    \caption{Clustering obtained by t-SNE on functional MRI of the Human Connectome Project. Each cluster represents a task.}
    \label{fig:task_separation}
\end{figure}

In this experiment, we show that the task being performed by an anonymized subject can be predicted with high accuracy. To demonstrate this result, we rely on t-SNE (Section \ref{sec:tsne}). We assume that the task labels of images from 50 subjects from the HCP are known. We stack the vectors of all 100 subjects into one matrix of dimensions $800 \times 64620$ (100 patients, each of who has one resting and seven task, for a total of 800 rows). Then, we use t-SNE to reduce the dimensions down to two. The resulting matrix has dimensions of $800 \times 2$. We call these two dimensional vectors as task-identifying signatures. Finally, we assign the task labels of the unknown data-points on the basis of their nearest neighbor with known task label. The result of this procedure is shown in Figure \ref{fig:task_separation}. The figure shows eight compact clusters -- one per task, separated cleanly in the 2-dimensional map. The only  exception is resting-state functional MRI (in pink). However, this is still extremely well clustered for real-world data. The objective function of t-SNE aims to respect pairwise distance between all pairs of points; hence this suggests that functional MRI of a subject is more similar to other subjects performing the same task than with the same subject performing different tasks. This observation is consistent with our previous experiment, where high accuracy is observed when both datasets corresponded to the same task. The task prediction accuracy using the nearest-neighbour approach over 100 iterations was 100\% for seven tasks and $99.01 \pm 0.52$ for the resting-state images. In instances where resting-state images were mis-classified, they were classified as gambling. This can be observed in Figure \ref{fig:task_separation} as well. 

\subsubsection{Prediction of performance during task}

In the previous experiment, we showed that a non-linear dimensionality reduction technique such as t-SNE is capable of finding the underlying cluster structure that allows us to predict what task is being performed by a subject. In this experiment, we predict how well they perform at the task. HCP provides task-performance metrics, which is a measure of effectiveness during the language processing, emotion processing, relational processing, and working memory tasks. This measure is expressed as a percentage accuracy of correct replies for the respective tasks. We show that our signatures can be used to predict the performance metrics of new subjects.

We divide the subjects into two sets, chosen at random -- train and test, of sizes 80 and 20 subjects, respectively. For each set of subjects, we compute the group matrices from the L-R images. Hence, the data matrix consists of vectorized correlation matrices of the subjects in train set for a particular task. The target vector consists of the performance metric for the same task. Similarly, a group matrix is constructed for the test set. Since we have the ground truth, i.e., the performance metrics for the test set, we can measure the goodness-of-fit for our model.

First, we compute leverage scores for the group matrix of train samples. We restrict the feature-space to the top leverage scores. We use SVM for regression, with the performance metric as the target. The coefficients learned from the  train-set are used to predict the scores in the test set. To avoid biases due to choices of train and test set, we repeat the experiment 1000 times.

The results are shown in Table \ref{tab:task_prediction} for both train and test. All the errors are within 4\% of the true value. We note that in our setup, we do not use timing information. This information encodes the exact time frames for the different blocks of stimuli within an experiment. Most studies also provide performance metrics within each time-block. The use of this additional data further improves prediction, and provides deeper insights that predict the neuronal response of individuals to particular sub-types of stimuli, such as math and story inputs, which is part of the language task.

\begin{table}
    \centering
    \begin{tabular}{|c|c|c|}
        \hline
         Task &  Train nRMSE (in \%) &Test nRMSE (in \%) \\  \hline
        Language & $0.33 \pm 0.11$ & $1.52 \pm 0.20$\\\hline
        Emotion & $0.28 \pm 0.07$ & $0.60 \pm 0.37$\\ \hline
        Relational & $0.44 \pm 0.04$ & $2.74 \pm 0.34$\\ \hline
        Working Memory & $0.57 \pm 0.12$ & $1.93 \pm 0.41$\\ \hline
    \end{tabular}
    \caption{Task-wise prediction error expressed as normalized root-mean-squared error.}
    \label{tab:task_prediction}
\end{table}

\subsubsection{Prediction of Identity of subjects with ADHD}

In this section, we extend the concept of brain signatures to subjects with neurological disorders such as Attention Deficit Hyperactive Disorder (ADHD). We will use the dataset from the ADHD-200 dataset. As before, we stack the vectorized connectomes of subjects into two matrices, one per session. Since we use the AAL2 atlas for the ADHD-200 dataset, each data-point now has 6670 features. 

\begin{figure}[h]
    \centering
    \includegraphics[scale=0.6]{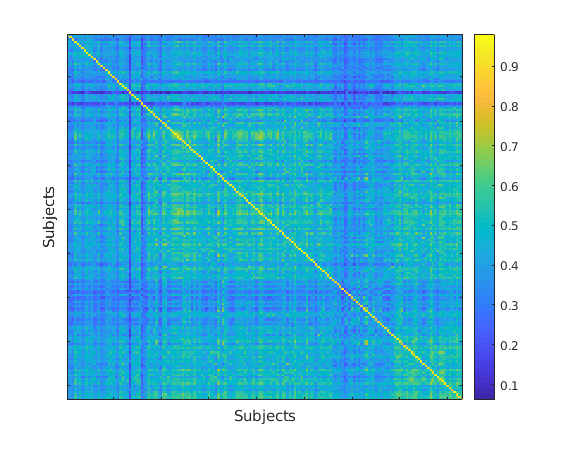}
    \caption{{\bf Correlation matrix of subjects with ADHD subtype 1 in the ADHD-200 dataset.} The strong values on the diagonal suggests strong inter-session similarity between scans of the same subject. The low off-diagonal values suggest a strong dissimilarity between scans of different subjects.}
    \label{fig:adhd_type1}
\end{figure}

\begin{figure}[h]
    \centering
    \includegraphics[scale=0.6]{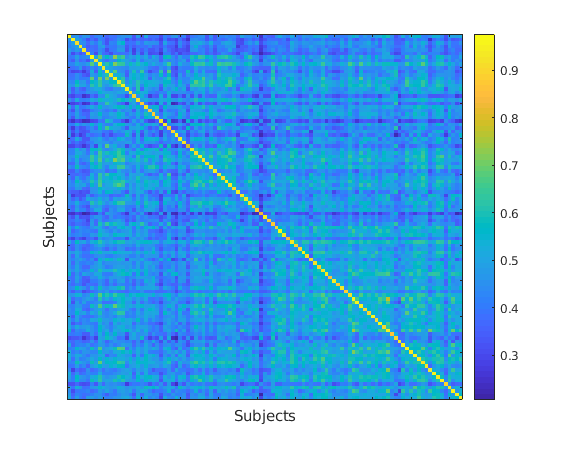}
    \caption{{\bf Correlation matrix of subjects with ADHD Subtype 3 in the ADHD-200 dataset.} Once again, we observe strong similarity between scans of the same subject.}
    \label{fig:adhd_type3}
\end{figure}

The stark difference between intra-subject similarity and inter-subject similarity for ADHD Subtypes 1 and 3 are shown in Figures \ref{fig:adhd_type1} and \ref{fig:adhd_type3}. We divide the subjects into train and test sets, and pick features that correspond to the highest leverage scores of the train matrix, as shown in Figure \ref{fig:my_pipeline}. When we restrict the feature space in the test subjects to the same features, we find the test accuracy to be $97.2 \pm 0.9$\%.

\begin{figure}
     \centering
     \includegraphics[scale=0.6]{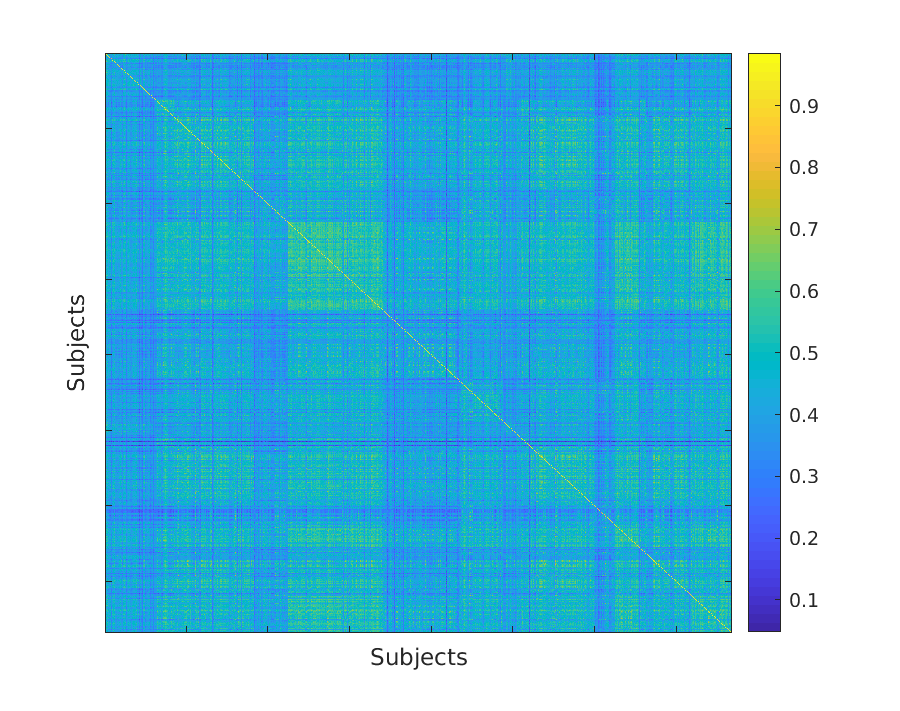}
     \caption{{\bf Correlation matrix of all subjects in the ADHD-200 dataset.} The leverage-score based sampling method is effective in finding signatures, even in datasets with both cases and controls.}
     \label{fig:adhd_all}
 \end{figure}

 This result is of significant interest for the following reasons. First, the acquisition protocol of ADHD-200 is different from HCP; also, the atlas used here is drawn differently from the one used in the previous experiments, and has different number of regions. Second, the subjects in this dataset are children with ADHD, whereas the HCP dataset is comprised of healthy young adults. Despite all of these differences, we find a consistent signature that can correctly identify subjects with high accuracy.  Finally, functional MRIs of patients exist in patient records outside of neuroimaging studies. Hence, identifiable data of such patients is often not stored with the same strict protocols, thereby increasing vulnerability to de-anonymization attacks. 
 Furthermore, even if the dataset is a combination of cases and controls, which is often the case, we can correctly match scans across datasets. In Figure \ref{fig:adhd_all}, we show that the strong values on the diagonal suggests a high prediction accuracy, which was observed to be $94.12 \pm 3.4$\%.
 
 \subsubsection{Effect of multi-site image acquisition on signatures}
 
 In previous experiments, we varied different parameters such as rest/ task, case/ control and parcellation schemes. However, in the most general case, different scans of a subject can be acquired at different sites/ hospitals. Therefore, the make and model of the instruments used to record the neuronal activity can be different for the two sessions. Hence, the accuracy of identification could be adversely affected.
 
 In this final experiment, we show that even in the case of multi-site acquitisition with different protocols, we can still predict the identity of an anonymized scan. We simulate a multi-site acquisition in the following manner: for each time-series signal in the second session of the Healthy Adult dataset by the HCP, we add Gaussian noise whose mean is equal to the mean of the original signal and whose variance is a fraction of the variance of the original signal. After this additional noising step, we compute the correlation matrices for the second session. Following this, the steps are identical to the previous setup (i.e., Figure \ref{fig:my_pipeline}). We compute the test accuracy for different noise values, and the results are summarized in Table \ref{tab:multisite}. For small values of noise, the prediction accuracy is still in excess of 90\%.
 
 We simulate the data for multi-site acquisition because of unavailability of large-scale publicly available datasets. However, this is a common occurence in practice, where a patient may have records at multiple hospitals. Furthermore, due to logistic reasons, a number of studies (including ADHD-200) are conducted in parallel at multiple institutions around the world. Hence, corrections for in-homogeneity in magnetic fields and other batch correction methods have been developed to enhance similarity across sites. It is for this reason that we believe that in our simulation, low values of noise variance reasonably simulate a multi-site acquisition.
 
 \begin{table}
 \centering
 \begin{tabular}{|c||c|c|}
  \hline
  \multirow{2}{*}{Noise Variance(in \%)} 
      & \multicolumn{2}{c|}{Identification Accuracy (in \%)}  \\             \cline{2-3}
  & HCP & ADHD-200 \\  \hline
  $10$ & $91.14 \pm 1.15$ & $96.33 \pm 0.79$\\ \hline
  $20$ & $86.71 \pm 2.44$ & $89.17 \pm 1.82$\\ \hline
  $30$ & $79.05 \pm 5.76$ & $84.10 \pm 4.43$\\ \hline
\end{tabular}
\caption{{\bf Summary of prediction accuracy for different noise levels.}}
     \label{tab:multisite}
 \end{table}
 
\section{Discussion \& Conclusion}
In this paper, we demonstrate a de-anonymization attack on neuroimaging data. Specifically, in functional MRIs, if one dataset of images is de-anonymized, we show that it is possible to de-anonymize any other datasets containing scans of the same subjects. This attack could potentially reveal other information contained in patient health records. We also show that it is possible to predict the task being performed, as well as the ability of a subject to perform specific tasks.

An effective defense to this type of an attack would require removal of the signature pattern without damaging the integrity of the image with respect to downstream analyses. In other words, every other property of the image, like the presence of a lesion or injury, change in relative sizes of different regions of the brain due to disease, etc. need to be maintained. To this end, our work presents a particularly important contribution, in that is clearly identifies a small signature that codes identity. By doing so, it provides a localized region where noise can be added to most effectively defend against such attacks. The impact of such noise added to targeted regions on downstream analyses is a function of the analyses, and must be assessed in specific application contexts.

\bibliographystyle{plainnat}
\bibliography{bibliography.bib}

\end{document}